\documentclass[aps,prl,reprint,amssymb,superscriptaddress]{revtex4-1}

\usepackage{amsmath,amssymb,color}
\usepackage{bm}
\usepackage{hyperref}
\usepackage{graphicx}

\newcommand{\updownarrows}{\uparrow\mathrel{\mspace{-1mu}}\downarrow}

\begin{document}

\title{Long-Range Phonon Spin Transport in Ferromagnet$-$Nonmagnetic Insulator Heterostructures}

\author{Andreas R\"{u}ckriegel}
\affiliation{Institute for Theoretical Physics and Center for Extreme Matter and Emergent Phenomena,
Utrecht University, Leuvenlaan 4, 3584 CE Utrecht, The Netherlands}

\author{Rembert A. Duine}
\affiliation{Institute for Theoretical Physics and Center for Extreme Matter and Emergent Phenomena,
Utrecht University, Leuvenlaan 4, 3584 CE Utrecht, The Netherlands}
\affiliation{Center for Quantum Spintronics, Department of Physics, Norwegian University of Science and Technology, 
NO-7491 Trondheim, Norway}
\affiliation{Department of Applied Physics, Eindhoven University of Technology,
P.O. Box 513, 5600 MB Eindhoven, The Netherlands}

\date{January 20, 2020}

\begin{abstract}
We investigate phonon spin transport in an insulating 
ferromagnet$-$nonmagnet$-$ferromagnet heterostructure.
We show that the magnetoelastic interaction between the spins and the phonons 
leads to nonlocal spin transfer between the magnets.
This transfer is mediated by a local phonon spin current 
and accompanied by a phonon spin accumulation.
The spin conductance depends nontrivially on the system size,
and decays over millimeter length scales for realistic material parameters,
far exceeding the decay lengths of magnonic spin currents.  
\end{abstract}

\maketitle

\paragraph{Introduction.}
One of the main goals of the field of spintronics is achieving long-range spin transport 
through electrical insulators \cite{Cornelissen2015,Lebrun2018}.
Up to now, the main focus of this research is on magnetic insulators,
in which the spin is carried by spin waves (or magnons) 
that are the elementary excitations of the magnetic order parameter.
However, it was shown recently that the magnetization dynamics in a ferromagnet can even
inject a spin current into an adjacent \emph{nonmagnetic} insulator \cite{Streib2018},
analogous to the spin pumping at the interface of a ferromagnet and a normal metal \cite{Tserkovnyak2005}. 
In this case, the spin is carried by transverse acoustic phonons with circular polarization \cite{Streib2018,Levine1962,Garanin2015,Holanda2018,Nakane2018}.
A long-range exchange coupling that was observed in a ferromagnet-semiconductor hybrid structure was similarly interpreted in terms of spin transfer by circularly polarized phonons \cite{Korenev2015}.
This raises the possibility of using phonon currents to transfer spin in (non-)magnetic insulators. 
Indeed, An {\it et al.} have found that phonons in nonmagnetic gadolinium-gallium garnet (GGG) 
mediate a coherent coupling between two yttrium-iron garnet (YIG) films that are half a millimeter apart \cite{An2019}.
In their experiment, An {\it et al.} coherently excited the ferromagnetic resonance (FMR) of a YIG film with a microwave field,
which affected the coherent FMR dynamics of a second YIG film separated from the first by a nonmagnetic GGG spacer.
This coupling is interpreted in terms of a phonon spin current,
with a propagation length that surpasses the analogous magnon propagation length by several orders of magnitude
because of the low acoustic damping in these materials. 
In view of possible spintronics applications, it would be desirable to drive the (phonon) spin current electrically 
instead of via a microwave field, 
e.g.~by exciting incoherent magnons via an electronic spin accumulation in a metallic lead \cite{Tserkovnyak2005,Cornelissen2015,Lebrun2018}.

\begin{figure}
\includegraphics[width=0.9\columnwidth]{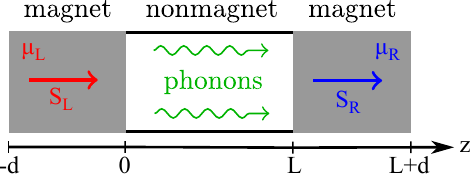}
\caption{ \label{fig:setup}
Heterostructure consisting of a nonmagnetic insulator of length $L$ 
sandwiched between two identical magnetic insulators of length $d$. 
The left and right spins interact via the phonons in the nonmagnetic insulator 
that may support a finite spin current between the magnets.
The magnon spin accumulations in the magnets are parametrized with magnon chemical potentials $\mu_{L/R}$.
These chemical potentials are generated by metallic leads (not shown) that are attached to the magnets.  
}
\end{figure}

In this work, 
we calculate the incoherent spin transport in an insulating ferromagnet$-$nonmagnet$-$ferromagnet heterostructure
that is driven by a difference in magnon spin accumulation between the magnets, 
as depicted in Fig.~\ref{fig:setup}.
We assume that the magnets are attached to metallic leads with fixed electronic spin accumulations $\mu_{L/R}$
that act as chemical potentials for the magnons \cite{Cornelissen2016}.
The metallic contacts are furthermore assumed to be small enough that they do not influence the phonon modes 
in the heterostructure.
We find that the magnetoelastic interactions between magnons and phonons 
lead to finite phonon spin accumulations in all three layers,
and show that a local phonon spin current mediates nonlocal spin transfer between the two ferromagnets 
over macroscopic distances. 
%


\paragraph{Theoretical description.}
We consider a heterostructure consisting of a nonmagnetic insulator of length $L$ 
sandwiched between two identical magnetic insulators of length $d$, see Fig.~\ref{fig:setup}.
We also assume that the magnetizations in both magnets are parallel to each other and to the phonon propagation direction,
which maximizes the magnetoelastic phonon pumping \cite{Streib2018}. 
In this setup, the spin of the left and right magnets interacts with the elastic lattice displacement field 
$ \bm{u}(\bm{r}) $ via the magnetoelastic Hamiltonian \cite{Gurevich1996}
\begin{equation} \label{eq:Hme}
{\cal H}_{\rm me} = \frac{1}{s^2} \sum_{X=L,R} \int_V \textrm{d}^3r \sum_{\alpha\beta} B_{\alpha\beta}
s_{X,\alpha}(\bm{r}) s_{X,\beta}(\bm{r}) \epsilon_{\alpha\beta}(\bm{r}) ,
\end{equation}
where $\alpha$ and $\beta$ run over the three spatial components $x$, $y$, $z$;
$s$ and $V$ are the saturation spin density and volume of both magnets,
$\bm{s}_{L/R}(\bm{r})$ is the local spin density,
and $\epsilon_{\alpha\beta}(\bm{r}) = \frac{1}{2} \left[ \partial u_\beta(\bm{r}) / \partial r_\alpha 
+ \partial u_\alpha(\bm{r}) / \partial r_\beta \right]$ is the linearized strain tensor.
$ B_{\alpha\beta} = \delta_{\alpha\beta} B_\parallel + (1-\delta_{\alpha\beta}) B_\bot $
are the magnetoelastic constants of an isotropic system.
Focusing on the uniform macrospin modes of both magnets,
we write 
$ \bm{s}_{L/R}(\bm{r}) \simeq \sqrt{ \frac{s}{2V} } ( \bm{e}_+ \psi_{L/R}^\dagger + \bm{e}_- \psi_{L/R} ) + \bm{e}_z s $,
where $\bm{e}_\pm = \bm{e}_x \pm i \bm{e}_y$ are circularly polarized transverse basis vectors,
and $\psi_{L/R}$ and $\psi_{L/R}^\dagger$ are destruction and creation operators for the macrospin magnon modes
that satisfy the bosonic commutation relations $[ \psi_X, \psi_{X'}^\dagger ] = \delta_{X,X'}$. 
In this case the magnetoelastic Hamiltonian (\ref{eq:Hme}) reduces to
%
%
%
%
\begin{align}
{\cal H}_{\rm me} = 
&
\frac{ B_\bot }{ \sqrt{2sd} } \biggl\{ 
\psi_L^\dagger \bm{e}_+ \cdot
\left[ \bm{u}(z=0) - \bm{u}(z=-d) \right]
\nonumber\\
&
+ \psi_R^\dagger \bm{e}_+ \cdot
\left[ \bm{u}(z=L+d) - \bm{u}(z=L) \right]
+ {\rm H.c.} \biggr\} .
\label{eq:Hme_macro}
\end{align}
Here,
$ \bm{u}(z) = \sqrt{\frac{d}{V}} \int\textrm{d}x\int \textrm{d}y \bm{u}(\bm{r}) $
is the normalized average of the displacement field on the device cross section,
which we assumed to be large compared to the thickness $d$ of the magnets.  
Note that the magnetoelastic coupling (\ref{eq:Hme_macro}) only enters via boundary conditions,
and that the macrospin magnons only couple to the circularly polarized transverse phonon fields 
$u_\pm(z) = u_\mp^\dagger(z) \equiv \bm{e}_\pm \cdot \bm{u}(z)$. 
These phonons actually carry the internal angular momentum, 
or phonon spin \cite{Streib2018,Levine1962,Garanin2015,Holanda2018,Nakane2018}, 
$ L_z = \int \textrm{d}z\, l_z(z) $,
with the phonon spin density
\begin{align}
l_z(z) 
&= \bm{e}_z \cdot \left\langle \bm{u}(z) \times \rho(z) \partial_t \bm{u}(z) \right\rangle
\\
&= \rho(z)\, \textrm{Im} \left\langle u_+^\dagger(z) \partial_t u_+(z) \right\rangle ,
\label{eq:lz}
\end{align}
where $\rho(z)$ is the local mass density.
We can thus interpret the magnetoelastic Hamiltonian (\ref{eq:Hme_macro}) in terms of spin transfer between the 
magnetic and elastic subsystems.

To investigate the spin transport in the heterostructure depicted in Fig.~\ref{fig:setup}, 
we employ semiclassical stochastic differential equations,
which is, however, equivalent to a fully quantum nonequilibrium Green's function approach
in the linear regime \cite{Zheng2017}.
For an elastically isotropic medium,
we therefore consider the equations of motion
\begin{subequations} \label{eq:udot}
\begin{align}
\partial_t u_+(z,t) 
&
= \rho^{-1}(z) \pi_+(z,t) , \\[.2cm]
\partial_t \pi_+(z,t) 
&= \partial_z \left[ \mu (z) \partial_z u_+(z,t) \right] -  2 \eta(z) \pi_+(z,t)
\nonumber\\[.1cm]
&
- \frac{ 2B_\bot }{ \sqrt{2sd} } 
\psi_L(t) \left[ \delta(z) - \delta(z+d) \right] 
\nonumber\\
&
- \frac{ 2B_\bot }{ \sqrt{2sd} }  
\psi_R(t) \left[ \delta(z-L-d) - \delta(z-L) \right] ,
\end{align}
\end{subequations}
where $\pi_+(z,t)$ is the momentum density conjugate to $u_+(z,t)$,
and $\rho(z)$, $\mu(z)$, and $\eta(z)$ are the local mass density, shear modulus and elastic damping constant.
For $-d<z<0$ and $L<z<L+d$, they are given by
$\rho(z) = \rho $, $\mu(z) = \rho c_\bot^2$, and $\eta(z) = \eta$ ,
whereas for $0<z<L$ they are
$\rho(z) = \tilde{\rho}$, $\mu(z) = \tilde{\rho} \tilde{c}_\bot^2$, and $\eta(z) = \tilde{\eta}$,
with the transverse sound velocities $c_\bot$ and $\tilde{c}_\bot$
in the magnetic and nonmagnetic insulators.
The macrospin magnon modes are governed by
\begin{subequations} \label{eq:psidot}
\begin{align}
& ( 1+i\alpha^G +i\alpha^{\rm sp} ) i\partial_t \psi_L(t) =
\left( \omega_{\textrm{FM}} + i\alpha^{\rm sp} \frac{\mu_L}{\hbar} \right) \psi_L(t) 
\nonumber\\
&
- h_L^G(t) - h_L^{\rm sp}(t)
+ \frac{ B_\bot }{ \hbar \sqrt{2sd} } \left[ u_+(0,t) - u_+(-d,t) \right],
\\
& ( 1+i\alpha^G +i\alpha^{\rm sp} ) i\partial_t \psi_R(t) =
\left( \omega_{\textrm{FM}} + i\alpha^{\rm sp} \frac{\mu_R}{\hbar} \right) \psi_R(t) 
\nonumber\\
&
- h_R^G(t) - h_R^{\rm sp}(t)
+ \frac{ B_\bot }{ \hbar \sqrt{2sd} } \left[ u_+(L+d,t) - u_+(L,t) \right] .
\end{align}
\end{subequations}
Here, 
$\omega_{\rm FM}$ is the FMR frequency,
$\alpha^G$ the Gilbert damping constant,
and $\alpha^{\rm sp} = g^{\updownarrows}/4\pi s d$ the Gilbert damping enhancement 
due to spin pumping to the metallic leads, with the spin-mixing conductance $g^{\updownarrows}$ \cite{Tserkovnyak2002}. 
Furthermore,
$ h_{L/R}^{G/{\rm sp}}(t) = (1/2\pi)\int\textrm{d}\omega e^{-i\omega t} h_{L/R}^{G/{\rm sp}}(\omega) $ 
are noise fields that satisfy the quantum fluctuation-dissipation theorems \cite{Zheng2017}
\begin{subequations}
\begin{align}
\left\langle h_{L/R}^G(\omega) h_{L/R}^{G*}(\omega') \right\rangle =
&
2\pi \delta(\omega-\omega') \alpha^G \omega
\coth\left( \frac{ \hbar \omega }{ 2 k_B T } \right) ,
\\
\left\langle h_{L/R}^{\rm sp}(\omega) h_{L/R}^{{\rm sp}*}(\omega') \right\rangle =
&
2\pi \delta(\omega-\omega') 
\frac{\alpha^{\rm sp}}{\hbar} \left( \hbar \omega - \mu_{L/R} \right)
\nonumber\\
&
\times
\coth\left( \frac{ \hbar \omega - \mu_{L/R} }{ 2 k_B T } \right) ,
\end{align}
\end{subequations}
with the magnon temperature $T$ and the left and right magnon chemical potentials $\mu_{L/R}$.
We do not consider a similar noise field for the phonons or temperature gradients because 
we focus on genuine long-range interaction between the two macrospin modes.
This interaction is mediated by the phonons that are driven by the magnon distributions.
In contrast, thermal phonons only add to the damping of the two macrospin modes 
when there are local temperature gradients across the magnet$-$nonmagnet interfaces.

The elastic equations of motion (\ref{eq:udot}) are supplemented by elastic continuity boundary conditions at the interfaces:
\begin{subequations} \label{eq:elastic_continuity}
\begin{align}
u_+ ( 0^+,t ) =& u_+ ( 0^-,t ) , \\
u_+ ( L^+,t ) =& u_+ ( L^-,t ) .
\end{align}
\end{subequations}
At the interfaces and boundaries, 
we furthermore impose momentum conservation, 
yielding
\begin{subequations} \label{eq:momentum_conservation}
\begin{align}
\mu(0^+) u_+'(0^+,t) - \mu(0^-) u_+'(0^-,t) - \frac{ 2B_\bot }{ \sqrt{2sd} } \psi_L(t) =& 0 ,
\\
\mu(L^+) u_+'(L^+,t) - \mu(L^-) u_+'(L^-,t) + \frac{ 2B_\bot }{ \sqrt{2sd} } \psi_R(t) =& 0 ,
\\
\mu(-d^+) u_+'(-d^+,t) + \frac{ 2B_\bot }{ \sqrt{2sd} } \psi_L(t) =& 0 ,
\\
- \mu(L+d^-) u_+'(L+d^-,t) - \frac{ 2B_\bot }{ \sqrt{2sd} } \psi_R(t) =& 0 ,
\end{align}
\end{subequations}
where $u_+'(z,t)=\partial_z u_+(z,t)$.

In a stationary state,
we may write
$ u_+(z,t) = (1/2\pi)\int\textrm{d}\omega e^{-i\omega t} u_+(z,\omega) $ and
$ \psi_{L/R}(t) = (1/2\pi)\int\textrm{d}\omega e^{-i\omega t} \psi_{L/R}(\omega) $;
then the solutions of the coupled magnetoelastic equations of motion (\ref{eq:udot}) and (\ref{eq:psidot}) 
are given by
\begin{widetext}
\begin{equation}
u_+(z,\omega) =
\begin{cases}
A(\omega) e^{ i k(\omega) ( z + d) } + B(\omega) e^{ - i k(\omega) z } , & -d < z < 0 , \\
C(\omega) e^{ i \tilde{k}(\omega) z } + D(\omega) e^{ i \tilde{k}(\omega) ( L - z ) } , & 0 < z < L , \\
E(\omega) e^{ i k(\omega) ( z - L ) } + F(\omega) e^{ i k(\omega) ( L + d - z ) } , & L < z < L + d ,
\end{cases}
\end{equation}
and
\begin{subequations}
\begin{align}
\psi_L(\omega) =
&
g_L(\omega) \left\{ - h_L^G(\omega) - h_L^{\rm sp}(\omega) + \frac{ B_\bot }{ \hbar \sqrt{2sd} }
\left( e^{ i k(\omega) d } - 1 \right) \left[ A(\omega) - B(\omega) \right] \right\} ,
\\
\psi_R(\omega) =
&
g_R(\omega) \left\{ - h_R^G(\omega) - h_R^{\rm sp}(\omega) + \frac{ B_\bot }{ \hbar \sqrt{2sd} }
\left( e^{ i k(\omega) d } - 1 \right) \left[ E(\omega) - F(\omega) \right] \right\} ,
\end{align}
\end{subequations}
\end{widetext}
with the phonon wave vectors 
$ c_\bot k(\omega) = \sqrt{ \omega^2 + 2 i \eta \omega } $ and 
$ \tilde{c}_\bot \tilde{k}(\omega) = \sqrt{ \omega^2 + 2 i \tilde{\eta} \omega } $,
and the noninteracting magnon Green's functions
\begin{equation}
g_{L/R}(\omega) = \frac{1}
{ \omega - \omega_{\textrm{FM}} + i \alpha^G \omega + i \alpha^{\rm sp} \left( \omega - \mu_{L/R} / \hbar \right) } .
%
\end{equation}
The elastic amplitudes $A(\omega),\ldots,F(\omega)$ are determined by the boundary conditions 
(\ref{eq:elastic_continuity}) and (\ref{eq:momentum_conservation});
as the explicit, analytical expressions for these amplitudes are rather involved, we omit them here.

\paragraph{Spin transport.}
Equipped with the full, analytical solution of the coupled magnetoelastic dynamics (\ref{eq:udot}) and (\ref{eq:psidot}),
we proceed to calculate various observables characterizing the spin transport through the heterostructure.
Throughout the remainder of this Letter,
we use parameters of YIG for the magnets
and GGG for the nonmagnet \cite{An2019,Gurevich1996,Cherepanov1993},
i.e., 
$s = 7.5/\textrm{nm}^{3}$,
$\rho = 5170\, \textrm{kg}/\textrm{m}^{3}$,
$c_\bot = 3843\, \textrm{m}/\textrm{s}$,
$B_\bot = 6.96\times 10^{5}\, \textrm{J}/\textrm{m}^{3}$, and
$ \alpha^G = 9 \times 10^{-5} $, 
as well as
$\tilde{\rho} = 7080\, \textrm{kg}/\textrm{m}^{3}$ and
$\tilde{c}_\bot = 3530\, \textrm{m}/\textrm{s}$.
The metallic leads are taken to be platinum, 
so that $ g^{\updownarrows} =  5\, {\rm nm}^{-2}$ \cite{Jia2011}.
Furthermore, 
we assume room temperature, 
$ T = 300\, {\rm K}$,
and fix the FMR frequency to
$\omega_{\rm FM}/ 2\pi = 5.49\, \textrm{GHz} $.
For these parameters, the 
wavelengths of phonons in the magnets and the nonmagnet are
$\lambda = 2\pi c_\bot / \omega_{\rm FM} = 700\, \textrm{nm}$ and
$\tilde{\lambda} = 2\pi \tilde{c}_\bot / \omega_{\rm FM} = 643\, \textrm{nm}$
respectively.
Lastly, the phonon damping is taken to be 
$ \eta / \omega_{\rm FM} = \tilde{\eta} / \omega_{\rm FM} = 6 \times 10^{-5} $,
corresponding to the damping rate measured in Ref.~\onlinecite{An2019}
for a heterostructure of the same materials at room temperature.

\begin{figure}
\includegraphics[width=0.9\columnwidth]{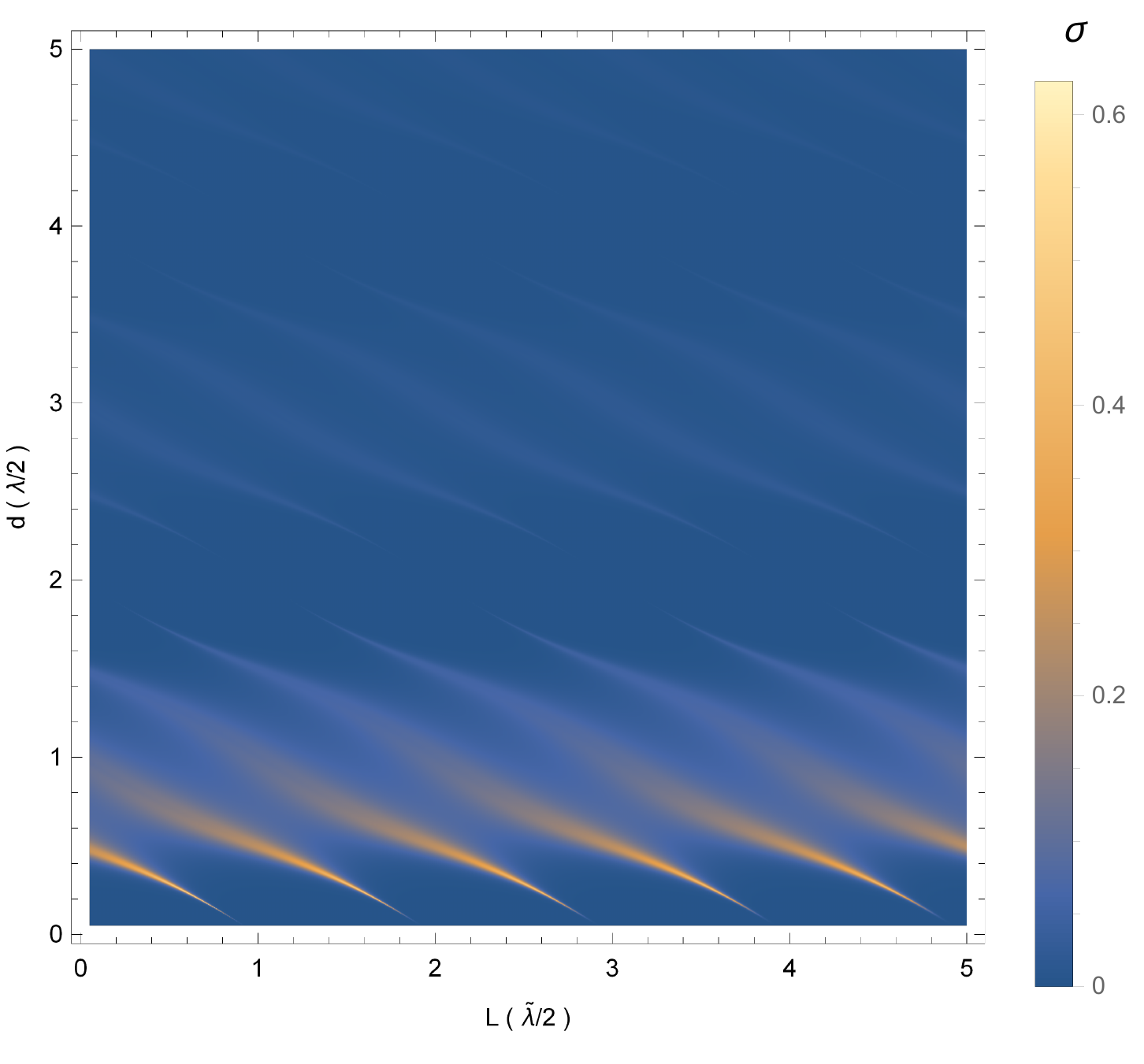}
\caption{ \label{fig:sigma}
Nonlocal spin conductance (\ref{eq:conductance}) as a function of 
the lengths $d$ and $L$ of the magnetic and nonmagnetic insulators,
for the parameters stated in the main text. 
The lengths are given in units of half wavelengths 
of phonons at the ferromagnetic resonance frequency in the respective material.
}
\end{figure}

The spin of the left and right macrospin is given by 
$ S_{R/L,z}(t) = \hbar \left[ sV + \frac{1}{2} - \left\langle | \psi_{R/L}(t) |^2 \right\rangle \right] $.
To obtain the spin current from the left to the right magnet that is detected in the right lead,
we consider the spin lost by the right macrospin to this lead: 
\begin{align}
\left( \partial_t S_{R,z} \right)^{\rm sp} =
&
2 \alpha^{\rm sp} \textrm{Re}
\left\langle
\psi_R^*(t) \left[ i\hbar \partial_t - \mu_R \right] \psi_R(t)
\right\rangle
\\
=
&
- I_{R} - I_{L\to R} .
\end{align}
Here, $I_R$ is a local contribution,
i.e., it only depends on the noise distribution of the right magnet itself.
On the other hand, 
$I_{L\to R}$ is a genuine nonlocal spin current from the left to the right magnet that is mediated by the phonons;
it is explicitly given by
\begin{align}
I_{L\to R} = 
&
\int \frac{\textrm{d}\omega}{2\pi}  {\cal T}_{L\to R}(\omega)
\nonumber\\
&
\times
\left[ 
f_B\left( \frac{ \hbar \omega - \mu_R } { k_B T } \right) - 
f_B\left( \frac{ \hbar \omega - \mu_L } { k_B T } \right)
\right] ,
\label{eq:current}
\end{align}
with the Bose function $f_B(x) = 1/\left( e^x - 1 \right)$
and the transmission function
\begin{align}
{\cal T}_{L\to R}(\omega) =
&
2 \frac{ \left(\alpha^{\rm sp}\right)^2 B_\bot^2 }{ \hbar^3 s d } 
\left( \hbar\omega - \mu_R\right) \left( \hbar\omega - \mu_L\right) 
\nonumber\\
&
\times
\left| g_R(\omega)
\left( e^{ i k(\omega) d } - 1 \right) 
\frac{ \partial \left[ E(\omega) - F(\omega) \right] }{ \partial h_L^{\rm sp}(\omega) } 
\right|^2 .
\label{eq:transmission}
\end{align}
For sufficiently small biasing, $\mu_{L/R} \ll \hbar \omega_{\rm FM}$,
we may further linearize the Bose functions.
Then the spin current (\ref{eq:current}) reduces to
$ I_{L\to R} = \sigma \left( \mu_R - \mu_L \right) $,
with the nonlocal spin conductance 
\begin{equation} \label{eq:conductance}
\sigma = \frac{1}{4 k_B T} \int \frac{\textrm{d}\omega}{2\pi}  
\frac{ {\cal T}_{L\to R}(\omega) }{ \sinh^2\left( \frac{ \hbar\omega }{ 2 k_B T } \right) } .
\end{equation}
An intensity plot of this conductance is shown in Fig.~\ref{fig:sigma}
as function of the lengths $d$ and $L$ of the magnetic and nonmagnetic insulators.
The conductance exhibits pronounced minima for  
$ d = 2 n \times \lambda/2 $, where $n=0,1,2,\ldots$
The reason for this is that the macrospin exerts forces of equal magnitude but opposite direction 
on the elastic field at the interfaces, see Eqs.~(\ref{eq:udot}).
Thus the excitation of phonons is favored when the length of the magnet is 
close to an odd number of phonon half-wavelengths
and suppressed when it is close to an even number.  
As function of the length $L$ of the nonmagnetic insulator
the conductance also shows a modulation, 
with local maxima along the lines 
$ L /( \tilde{\lambda}/2 ) + 2 d /( \lambda/2 ) = m$, 
where $m = 1, 2, \ldots$
This corresponds to standing waves for the whole heterostructure.
There is a slight deviation of the maxima from these straight lines because the hybridization of magnon and phonon modes leads to an anticrossing that shifts the phonon frequency away from the FMR frequency. 
The decay of the conductance for increasing size $d$ of the magnets is explained by the decay of both
the macrospin-phonon coupling [see Eq.~(\ref{eq:Hme_macro})], 
and the Gilbert-damping enhancement $\alpha^{\rm sp}\propto 1/d$
that couples the magnets to the leads, 
as a function of the size of the magnets.

\begin{figure}
\includegraphics[width=0.9\columnwidth]{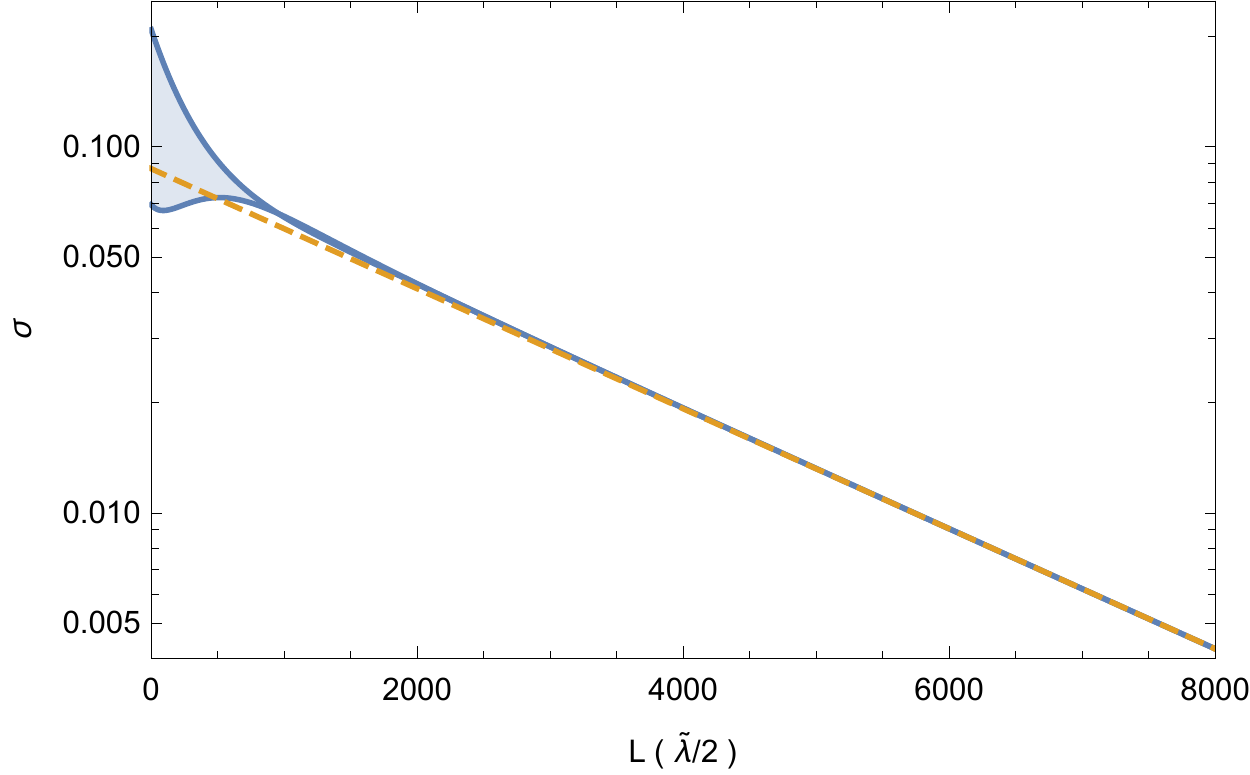}
\caption{ \label{fig:sigmaL}
Decay of the spin conductance (\ref{eq:conductance}) as a function of the length $L$
of the nonmagnetic insulator in units of half-wavelengths,
for $d=2\lambda/7 = 200\,{\rm nm}$
and the parameters stated in the main text.
Solid lines denote the envelope of the conductance curve;
in the shaded area the conductance oscillates rapidly between the minimal and maximal values given by the envelope,
see also Fig.~\ref{fig:sigma}.
The dashed line is an exponential fit with decay length 
$\approx 2650 \times \tilde{\lambda}/2 \approx 0.85\,{\rm mm}$.
}
\end{figure}

The behavior of the conductance for large nonmagnetic insulators is displayed in Fig.~\ref{fig:sigmaL},
on a logarithmic scale.
Similar to magnon conductances in magnetic insulators \cite{Cornelissen2015,Cornelissen2016},
the conductance exhibits a power law decay for small $L$,
and eventually decays exponentially for large $L$.
However, note that the crossover occurs at 
$L \approx 1000 \times \tilde{\lambda}/2 \approx 0.3\,{\rm mm}$ for our parameters,
while the characteristic decay length in the exponential regime is $ 0.85\,{\rm mm}$.
Both of these length scales are almost two orders of magnitude larger than the analogous 
length scales of magnon spin currents in YIG \cite{Cornelissen2015,Cornelissen2016}.
Another difference to magnons is that
because of the constructive interference for standing waves in the heterostructure (see Fig.~\ref{fig:sigma}),
the phonon conductance rapidly oscillates as a function of distance in the power-law decay regime.

\begin{figure}
\includegraphics[width=\columnwidth]{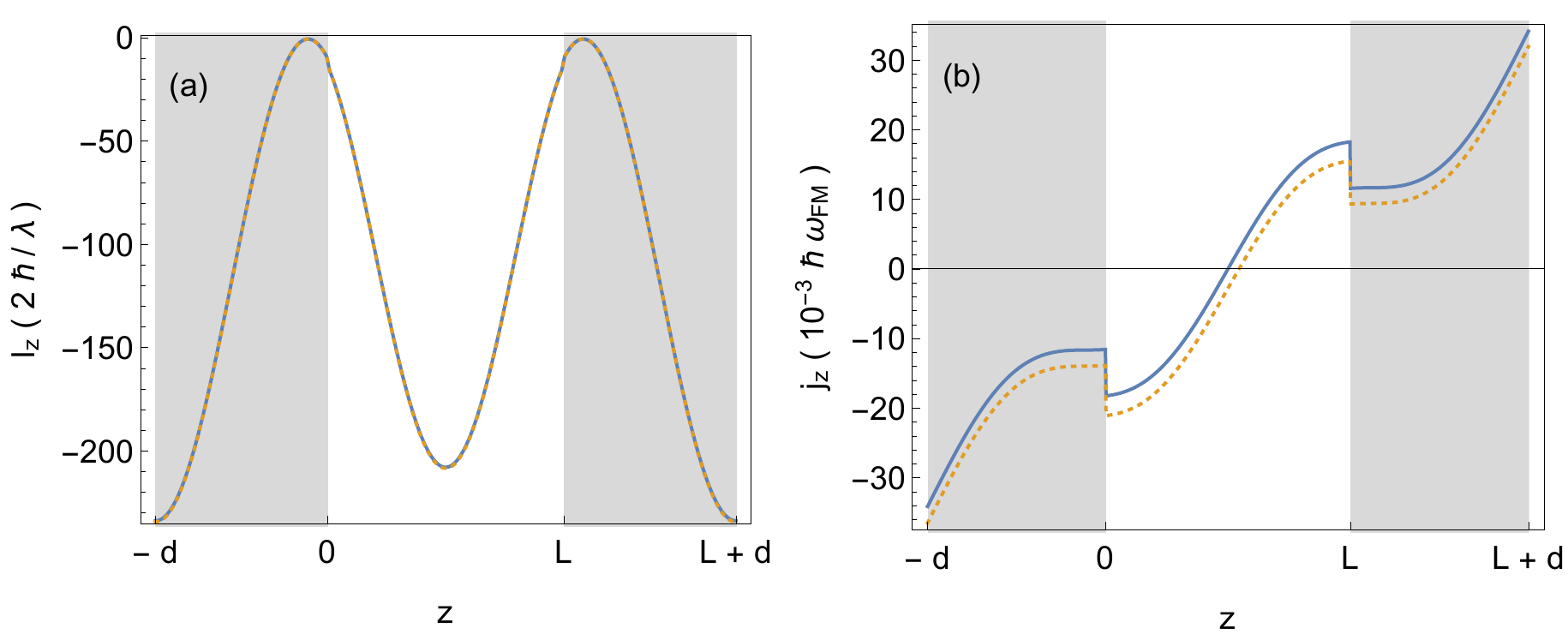}
\caption{ \label{fig:densities}
(a) Phonon spin and (b) phonon spin current densities for $\mu_L=0=\mu_R$ (solid lines),
and $\mu_L=0.01 \times \hbar \omega_{\rm FM}$ and $\mu_R=0$ (dashed lines)
for the parameters stated in the main text.
The system size is set to $d=2\lambda/7=200\,{\rm nm}$ and $L= 3\tilde{\lambda}/7=275.6\,{\rm nm}$.
The shaded region denotes the magnets.
The small jumps in $l_z$ at the interfaces are due to the different mass densities of the magnets
and the nonmagnet.
On the other hand, 
the jumps in $j_z$ are not only caused by the change of shear modulus 
but also by the magnetoelastic coupling, see Eqs.~(\ref{eq:momentum_conservation})
For comparison, the macrospin magnon spin densities in both magnets are $\approx -600\times 2\hbar/\lambda$.
}
\end{figure}

To further substantiate our claim of long-range phonon spin transport,
we note that away from the interfaces,
the phonon spin density (\ref{eq:lz}) satisfies a continuity equation:
\begin{align}
\partial_t l_z(z,t) +  \partial_z j_z(z,t)
&= - 2 \eta(z) l_z(z,t) ,
\end{align}
where
\begin{equation} \label{eq:jz}
j_z(z,t) = - \textrm{Im} \left\langle u_+^*(z,t) \partial_z \left[ \mu(z) u_+(z,t) \right] \right\rangle 
\end{equation}
is the local phonon spin current density.
Both the phonon spin and spin current densities are shown in Fig.~\ref{fig:densities}.
Because of the spin-transfer via the magnetoelastic interaction 
there is a phonon spin accumulation in all three layers, see Fig.~\ref{fig:densities} (a).
For the same reason the phonon spin current shown in Fig.~\ref{fig:densities} (b) is finite
even in the absence of biasing, $\mu_L=0=\mu_R$.
However, this current is symmetric around zero and consequently does not lead to a net spin transfer
between the magnets, in contrast to the biased setup with $\mu_L-\mu_R\neq 0$.
Note also that for the parameters shown in Fig.~\ref{fig:densities}
that correspond to a standing wave, i.e., a maximum of the conductance,
the phonon spin density is of the same order as the spin density of the macrospin magnons in the magnets.
For  a minimum of the conductance or in the exponentially decaying long-range regime on the other hand, 
the phonon spin density is generally at least two orders of magnitude smaller than the magnon spin.

Lastly, one can show that the total spin $J_z=L_z+S_{L,z}+S_{R,z}$ satisfies the equation of motion
\begin{align}
\partial_t J_z(t)
=
&
-2 \int\textrm{d}z \eta(z) l_z(z,t)  
\nonumber\\
& 
- 2\hbar \sum_{X=L,R} \textrm{Im} 
\left\langle \psi_X^*(t) \left[ \alpha^G \partial_t \psi_X(t) - h_X^G(t) \right] \right\rangle 
\nonumber\\
&
- 2\hbar \sum_{X=L,R} \textrm{Im} 
\left\langle \psi_X^*(t) \left[ \alpha^{\rm sp} \partial_t \psi_X(t) - h_X^{\rm sp}(t) \right] \right\rangle 
\nonumber\\
&
- 2 \alpha^{\rm sp} \sum_{X=L/R} \mu_X \langle \left| \psi_X(t) \right|^2 \rangle .
\end{align}
Therefore the total spin is conserved in the absence of dissipation ($\eta=0=\alpha^{G/{\rm sp}}$),
and the damping and noise terms model the loss of spin to the environment
consisting of nonuniform magnons, electronic leads, thermal phonons and the rigid-body dynamics of the lattice. 
In consequence, the nonlocal spin current (\ref{eq:current}) between the two magnets must be mediated by
the local phonon spin current density (\ref{eq:jz}).

\paragraph{Discussion and conclusions.}
Utilizing the phonon degree of freedom in (non-)magnetic insulators provides a novel route for long-range spin transport.
We have shown that
there is a finite phonon spin accumulation as well as a finite phonon spin current
in an insulating magnet$-$nonmagnet$-$magnet heterostructure
driven by the magnon distributions in the magnets. 
If those magnon distributions are not in equilibrium with each other,
there is a net spin current mediated by the phonons.
For realistic material parameters,
we have found that this nonlocal spin current decays over millimeter length scales 
that are significantly larger than the decay lengths of magnonic spin currents in magnetic insulators.  
A direct comparison of the magnitude of the phonon and magnon spin currents is less straightforward;
however, since the magnon spin current is carried by a continuum of thermal magnons 
while the phonon spin current is driven by the single FMR mode,
we expect this phonon spin current to be small compared to the magnon spin current at room temperature. 
Experimentally,
the phonon spin current is detectable electrically via the inverse spin Hall effect in the metallic leads 
\cite{Cornelissen2015,Cornelissen2016,Tserkovnyak2005}.	
The predicted phonon spin accumulation should be observable with Brillouin light scattering \cite{Holanda2018}.

Because the spin transfer from the magnons to the phonons depends on a coherent magnon-phonon interconversion process
at the interfaces,
the nonlocal spin transport is particularly sensitive to the length of the magnets,
and to a lesser extent also to the length of the nonmagnet.
In particular, spin transport is almost completely prohibited when the length of the magnets
corresponds to an integer multiple of the phonon wavelength at the ferromagnetic resonance frequency.
This makes it possible to switch between a spin-conducting and a spin-nonconducting state 
by changing the ferromagnetic resonance frequency of the magnets, e.g. via an external magnetic field.

While we have shown that long-range spin transport via acoustic phonons is possible,
additional research is required to understand the effect of the phonon spin and angular momentum conservation 
in spin Seebeck experiments \cite{Uchida2010},
and to understand the relaxation of the phonon spin beyond phenomenological models.




\paragraph{Acknowledgments.}
We acknowledge useful discussions with Simon Streib and Gerrit E.~W.~Bauer.
This work is supported by the European Research Council via Consolidator Grant No.~725509 SPINBEYOND. 
R.D. is a member of the D-ITP consortium, a program of the Netherlands Organisation for Scientific Research (NWO) 
that is funded by the Dutch Ministry of Education, Culture and Science (OCW).
This research was supported in part by the National Science Foundation under Grant No.~NSF PHY-1748958.





%

%

\end{document}